\documentstyle[epsfig,12pt]{article}
\newcommand{\beq}{\begin{eqnarray}}
\newcommand{\eeq}{\end{eqnarray}}
\begin{document}
\baselineskip=24pt
\parindent 5mm
\vspace{-20ex}
\begin{center}
{\Large \bf Prediction of Cosmological Constant $\Lambda$ In
Veneziano Ghost Theory of QCD$^{*}$}\footnote {$^{*}$ This work was
supported in part by National Natural Science Foundation of China
(10647002), Guangxi Science Foundation for Young Researchers under
contract No. 0991009, and Guangxi Education Department with grant
No.200807MS112, Department of Science and Technology of Guangxi
under funds No. 2011GXNSFA018140, Department of Guangxi Education
for the Excellent Scholars of Higher Education, 2011-54, Doctoral
Science Foundation of Guangxi University of Technology, 11Z16, and
in part by the Pittsburgh Foundation}
\end{center}

\centerline{\bf Zhou Li-juan$^{1}$,~ Ma Wei-xing$^{2}$, Leonard S.
Kisslinger$^{3}$} \centerline{$^1$ Department of information and
computing science, Guangxi University of Technology,} \centerline{
Liu Zhou, 545006, Guangxi, China} \centerline{$^2$ Institute of high
energy physics, Chinese Academy of Sciences, Beijing, 100049, China}
\centerline{$^3$ Department of physics, Carnegie-Mellon University,
Pittsburgh, PA 15213, USA.}
\begin{center}
{\Large\bf Abstract}
\end{center}
Based on the Veneziano ghost theory of QCD, we estimate the
cosmological constant $\Lambda$, which is related to the vacuum
energy density, $\rho_{\Lambda}$, by $\Lambda = 8\pi G \rho_{\Lambda}$. 
In the recent Veneziano ghost theory
$\rho_{\Lambda}$ is given by the absolute value of the product of
the local quark condensate and quark current mass:$\rho_{\Lambda} =
\frac{2N_{f}H}{m_{\eta'}}c |m_{q}<0|:\bar{q}q:|0>|$. By solving
Dyson-Schwinger Equations for a dressed quark propagator, we found
the local quark condensate $<0|:\bar{q}q:|0> \simeq -(235 MeV)^{3}$,
the generally accepted value.  The quark current mass is 
$m_{q} \simeq$ 4.0 Mev. This gives the same result for 
$\rho_{\Lambda}$ as found by previous authors, which is somewhat larger 
than the observed value. However,
when we make use of the nonlocal quark condensate,
$<0|:\bar{q}(x)q(0):|0>= g(x)<0|:\bar{q}q:|0>$, with g(x) estimated
from our previous work, we find  $\Lambda$ is in a good agreement
with the observations.

\vspace{0.2cm} {\large \bf Key Words}: Cosmological constant
$\Lambda$, Veneziano Ghost theory of QCD, Local quark vacuum
condensate, Nonlocal quark condensate, Quantum Chromodynamics-QCD.

{\large \bf PACS} Number(s): 98.80.-k, 95.36.+x, 95.30.Sf, 12.38.Lg.

\newpage
\section{The Cosmological Constant $\Lambda$ and 
the QCD Veneziano Ghost Theory }

The starting point of most cosmological study is Albert Einstein's
Equations, which is a set of ten equations in Einstein's theory of
general relativity. The original Einstein field equations can be
written as the form$^{[1]}$
\begin{eqnarray}
R_{\mu\nu}-\frac{1}{2} R g_{\mu\nu}=8\pi G T_{\mu\nu}
\end{eqnarray}
in units of $\hbar = c = 1$, where G is the gravitational constant
($G = 6.7087(10)\times 10^{-39}GeV^{-2}$, sometime called Newton's
constant), $R_{\mu\nu}(\mu,\nu = 0, \cdots, 3)$ is the Ricci
tensor, R is the trace of Ricci tensor (it is like the radius of
curvature of space-time), $g_{\mu\nu}(x)$ represents the metric
tensor, which is a function of position $x$ in spacetime.
$T^{\mu\nu}$ is the energy-momentum tensor, which describes the
distribution of matter and energy. Eq.(1) describes a non-static
universe. However, Einstein believed, at that time, that our
universe should be static. In order to get a static universe, in
1917 Einstein introduced a new term, $\Lambda g_{\mu\nu}$, in Eq.(1)
to balance the attractive force of gravity, giving his
modified equation
\begin{eqnarray}
\label{lambda}
R_{\mu\nu}-\frac{1}{2} R g_{\mu\nu} + \Lambda g_{\mu\nu} = 8\pi G
T_{\mu\nu}.
\end{eqnarray}
The $\Lambda$ in Eq.(2) is the so-called cosmological constant, which
is a dimensional parameter with units of $(length)^{-2}$. Indeed,
Eq.(2) allows a static universe$^{[2]}$, called Einstein's
universe, which is one of the solution$^{[3]}$ of Friedmann's simplified
form of Einstein's equation with a $\Lambda$ term.
However, almost one hundred years ago the observations of redshifts
of galaxies led to Hubbles Law$^{[4]}$ and the interpretation that
the universe is expanding. This led Einstein to declare his static 
cosmological model, and especially the introduction of the $\Lambda$ 
term to his original field equation theory, his "biggest blunder".

  Note that the term $\Lambda g_{\mu\nu}$ in Eq.(\ref{lambda}) corresponds
to adding a vacuum term to $T_{\mu\nu}$,
\beq
\label{rhovac}
 T_{\mu\nu}(vac)&=& \rho_{\Lambda} g_{\mu\nu}.
\eeq
Therefore, the cosmological constant $\Lambda$ is related to
the vacuum energy density, $\rho_{\Lambda}$ by$^{[3]}$  
\begin{eqnarray}
\label{lambda2}
\Lambda = 8\pi G \rho_{\Lambda}.
\end{eqnarray}

  The vacuum energy density, called dark energy density, and a model
with $\Lambda$ representing dark energy were reintroduced about three decades
ago. See Ref.[5] for a review of the physics and cosmology of $\Lambda$,
with references to the many models that have been published.
To explain our uniform and flat universe via inflation a cosmological
constant was added to the Friedmann equation$^{[6]}$. From studies of 
radiation from the early universe, the Cosmic Microwave Background Radiation
(CMBR), by a number of projects, including WMAP$^{[7]}$, the inflation scenerio
was verified, and it was shown that about $73\%$ of the
total energy in the universe is dark energy. As clearly shown by Friedmann's 
equation with a cosmological constant, dark energy corresponds to negative
pressure, or anti-gravity. This was confirmed by studies of distant type 1a
supernovae$^{[8-9]}$, which showed an acceleration of the expansion of the
universe, and was consistent with dark energy being  $73\%$ of the energy
in the universe. Also, dark energy causes distant galaxies to
accelerate away from us, in contrast to the tendency of ordinary
forms of energy to slow down the recession of distant objects. See Ref.[5]
for other of the many references to CMBR, supernovae, galaxy and other
studies of dark energy

The existence of a non-zero vacuum energy would, in
principle, have an effect on gravitational physics on all scales.
The value of $\Lambda$ in our present universe is not well known, and it
is an empirical issue which will ultimately be settled by
observation. A precise determination of this number ($\Lambda$) or
$\rho_{\Lambda}$ will be one of the primary goals of observational cosmology 
in the near future. Recently the possiblity of determining the
cosmological constant by observations has been discussed$^{[10]}$.

A major outstanding problem is that most quantum field theories
predict a huge cosmological constant $\Lambda$ from the energy of
the quantum vacuum. This conclusion also follows from dimensional
analysis and effective field theory down to the Planck scale, by
which we would expect a cosmological constant of the order of
$M_{pl}^{4}$ ($M_{pl}$ is the Planck mass with $M_{pl}=
G^{-1/2}=1.22 \times 10^{19}GeV$. The Planck energy is thought to be
the energy where conventional physical theories break down and a new
theory of quantum gravity is required ). We know that the measured
value is on the order of $10^{-35}s^{-2}$,or $10^{-47}GeV^{4}$, or
$10^{-29}g/cm^{3}$, or about $10^{-120}$ in reduced planck units
($M_{pl}$). That is, there is a large difference between the 
magnitude of the vacuum energy expected from zero -point
fluctuations and scalar potential, $\rho^{theory}_{\Lambda} = 2
\times 10^{110} erg / cm^{3}$, and the observed value,
$\rho^{observe}_{\Lambda} = 2 \times 10^{-10} erg /cm^{3}$, a
discrepancy of a factor of $10^{120}$. This is the largest
discrepancy - the worst theoretical prediction in the history of
physics. At the same time, some supersymmetric theories require a
cosmological constant that is exactly zero. Therefore, we face a big
difficulty in understanding the observational
$\rho^{observe}_{\Lambda}$. This problem has been referred to as the
longstanding cosmological constant problem.

 Vacuum energy is predicted to be created in cosmological phase transitions.
In the standard model of particle physics with the temperature (T) of the
universe as a function of time (t), there are two important phase transitions.
At t $\simeq 10^{-11}$ seconds, with T $\simeq$ 140 GeV the universe undergoes
the electroweak phase transition (EWPT), with the vacuum expectation value
of the Higgs field, $\langle 0 \mid :\Phi^{Higgs} :\mid 0\rangle$, 
going from zero to a finite value corresponding to a
Higgs mass $\simeq$ 140 GeV. At t $\simeq 10^{-5}$ seconds, with T $\simeq$
150 MeV, the universe undergoes the QCD phase transition (QCDPT), when a
universe consisting of a dense quark-gluon plasma becomes our current universe
with hadrons. The latent heat for this phase transition is the quark
condensate, $\langle 0 \mid :\bar{q} q :\mid 0\rangle$, also a vacuum energy,
which is an essential part of the present work. 

  First we review the work of F. R. Urban, A. R. Zhitnitsky
$^{[11-12]}$, which is based on the QCD Veneziano ghost theory$^{[13-16]}$
In this model the cosmological vacuum energy density
$\rho_{\Lambda}$ can be expressed in terms of QCD parameters for
$N_{f}=2$ light flavors as follows$^{[10-11]}$
\begin{eqnarray}
\label{rholambda}
\rho_{\Lambda} = c \frac{2HN_{f}}{m_{\eta'}}\mid m_{q} \langle 0\mid
: \bar{q}(0)q(0) : \mid 0 \rangle \mid,
\end{eqnarray}
where  $ m_{q}$ is the current quark mass and $c = c_{QCD}\times c_{grav.}$. 
The first factor $c_{QCD}$ is a
dimensionless coefficient with value of $c_{QCD} \simeq
1$$^{[10-11]}$, which is entirely of QCD origin and is related to
the definition of QCD on a specific finite compact manifold such as
a torus, $ \rho_{\Lambda}\simeq c_{QCD} \frac{2N_{f}\mid m_{q}
\langle \bar{q}q \rangle\mid}{L m_{\eta'}}$ with $L$ being the size
of the manifold and $m_{\eta'}$ the mass of $\eta'$ meson. A precise
computation of $c_{QCD}$ has been calculated in a conventional
lattice QCD approach by studying corrections of order $1/s$ to the
vacuum energy $^{[10-11]}$. Note that $c_{QCD}$ depends on
the manifold where the theory is defined. The second factor
$c_{grav.}$ has a purely gravitational origin and is defined as the
relation between the size $L$ of the manifold we live in, and the
Hubble constant $H$, $L = (c_{grav.}H_{0})^{-1}$. One
can define this size of the manifold as $ L \simeq 17 H^{-1}_{0}$
where $H_{0} = 2.1\times 10^{-42}\times h GeV$ and $h = 0.71$
($H_{0}$, Hubble constant today). Therefore, one can explicitly
obtain an estimate for the linear length $L$ of the torus, and then
obtain the value of $c_{grav.}$ with $c_{grav.} = 0.0588$.

  In Section 2 we briefly review our previous calculation of the
quark condensate$^{[17]}$ using Dyson-Schwinger equations (DSEs)$^{[18-19]}$,
and discuss the quark current mass $m_{q}$, which are needed to
calculate $\rho_{\Lambda}$, as shown in Eq(\ref{rholambda}).
Since our values for the local quark condensate $\langle 0\mid
: \bar{q}(0)q(0) : \mid 0 \rangle$ and the current quark mass are
approximately the same as in Ref. [10,11] we find the same value
for $\rho_{\Lambda}$ as in that work,
with a factor 6 discrepancy when compared to the observed vacuum
energy density. In Sect. 3
we use a nonlocal quark condensate, based on earlier research, and
find good agreement between $\rho_{\Lambda}^{nonlocal\; theory}$ and
$\rho_{\Lambda}^{observed}$. Finally, we give our Summary and
concluding remarks in Sect.4.

\section{Local quark condensate, current quark mass,  
$\rho_{\Lambda}$}

   In this section we review our previous work on the quark condensate,
the current quark mass, and the resulting value for the cosmological 
constant/vacuum energy density.

\subsection{The local quark condensate}

The quark propagator is defined by
\begin{eqnarray}
\label{qprop}
S_{q}^{ab}(x) &=& \langle 0 \mid T[q^a(x)\bar{q}^b(0)] \mid 0 \rangle \; ,
\end{eqnarray}
where $q^a(x)$ ($q^{b}(x)$) is a quark field with color $a$ ($b$),
and $T$ is the time-ordering operator. The nonperturbative part of
the quark propagator is given by
\begin{eqnarray}
\label{NP} && S_{q}^{NP}(x) =-\frac{1}{12} [ \langle 0 \mid : \bar
{q}(x){q}(0): \mid 0 \rangle + x_{\mu} \langle 0 \mid :
\bar{q}(x)\gamma^{\mu} q(0): \mid 0 \rangle ] .
\end{eqnarray}
\newpage

For short distances, the Taylor expansion of the scalar
part,$\langle 0 \mid : \bar{q}(x) q(0):\mid 0 \rangle $, of
$S_{q}^{NP}(x)$ can be written as ( see, e.g., Refs.[17,20] )

\begin{eqnarray}
\label{qcond}
&&\langle 0 \mid : \bar q(x){q}(0): \mid 0 \rangle=\langle 0 \mid :
\bar q(0){q}(0): \mid 0 \rangle \nonumber \\
&&-\frac{x^{2}}{4}\langle 0 \mid : \bar q(0)[ig_{s}\sigma
G(0)]{q}(0): \mid 0 \rangle +\cdots .
\end{eqnarray}
In Eq.(\ref{qcond}) the vacuum expectation values in the expansion are the
local quark condensate, the quark-gluon mixed condensate, and so forth.

   The Dyson-Schwinger Equations$^{[18,19]}$ were used to
derive the local quark condensate in Ref.[17]. See this reference for
details and a discussion of approximations.
Note that as shown in Eq.(\ref{qcond}), the quark - gluon mixed condensate
provides the small-x dependence of the nonlocal
$\langle 0\mid : \bar{q}(x)q(0):\mid 0\rangle$ quark condensate. However,
for the present work this small-x expansion is not useful, and we shall
use a known expression for the nonlocality, described below. Therefore
we only give the results for the local quark condensate. Also note that
the vacuum condensates can act as a medium$^{[21-22]}$, which influences the 
properties of particles propagating through it.

Using the solutions of DSEs with three different sets of the quark-quark 
interaction parameters (see Ref.[17]) leads to our theoretical predictions 
for the local quark vacuum condensate listed in Table 1.
\begin{center}
{Table 1. Predictions of local quark condensate in QCD
vacuum, $\langle 0 \mid :\bar{q}q:\mid 0 \rangle_{\mu}^{f}$ with $f$
standing for quark flavor and $\mu$ denotes renormalization point,
$\mu^2$=10 GeV$^2$}.

\vspace{0.4cm}
\begin{tabular}{|c|c|c|}\hline
Set no. of quark interactions & $\langle 0\mid :\bar{q}q:\mid
0\rangle_{u,d}$ for $u$ and $d$ quarks \\
\hline Set 1 & $-0.0130 (\rm GeV)^{3} \sim - (235 MeV)^{3}$\\
\hline Set 2 & $-0.0078 (\rm GeV)^{3} \sim - (198 MeV)^{3}$\\
\hline Set 3 & $-0.0027 (\rm GeV)^{3} \sim - (139 MeV)^{3}$\\
\hline
\end{tabular}
\end{center}
Set 1 results are consistent with many other calculations, such as
QCD sum rules$^{[23,24,25]}$, Lattice QCD$^{[26,27,28]}$ and Instanton
model predictions$^{[29,30,31]}$. These numerical results will be used
to calculate $\Lambda$/$\rho_{\Lambda}$ in the subsection 2.3 below.

\subsection{The current mass of light quarks }

As we have seen from  Eq.(\ref{rholambda})  to predict
$\Lambda$ we need to know the basic quark current mass $m_{q}$.
Since one cannot produce a beam of quarks, it is difficult to
determine the quark masses. Using various models the effective
quark masses have been estimated, but we need the current quark masses of
the light u and d quark. Estimates of these masses and references can be
foud in the Particle Data Physics booklet$^{[32]}$. They are
\beq
\label{mud}
             &&\;1.7 < \;m_u \;< \; 3.3 {\rm \;MeV} \nonumber \\
              &&\;4.1 < \;m_d \;<\; 5.8 {\rm \;MeV} e \; .
\eeq

  From this we estimate that the current quark mass is
\beq
\label{mq}
               m_q &\simeq & 4.0 {\rm \;MeV} \; ,
\eeq

\subsection{Cosmological constant $\Lambda$ with
$\langle 0\mid : \bar{q}(0)q(0):\mid 0\rangle$ and $m_q$}

  From Eq.(\ref{lambda}), $\Lambda = 8\pi G \rho_{\Lambda}$,
the vacuum energy density, while $\rho_{\Lambda}$, is given in 
Eq(\ref{rholambda}) as
\begin{eqnarray}
\label{rho}
     \rho_{\Lambda} =  c \frac{2HN_{f}}{m_{\eta'}} m_q
|\langle 0\mid : \bar{q}(0)q(0):\mid 0\rangle| \;.
\end{eqnarray}
Since our values for $m_q$ and $\langle 0\mid : \bar{q}(0)q(0):\mid
0\rangle$ are the standard ones, we find the same value for
$\rho_{\Lambda}$ as in Ref.[11]
\begin{eqnarray}
\label{rhovalue}
     \rho_{\Lambda}^{theory} \simeq (3.6 \times 10^{-3}eV)^4 \;,
\end{eqnarray}
while the value observed$^{[33]}$ is
\begin{eqnarray}
\label{rhovalueobs}
     \rho_{\Lambda}^{observed} \simeq (2.3 \times 10^{-3}eV)^4 \;.
\end{eqnarray}

Although the theoretical and observed values are similar, they still
differ by \\
 $ \rho_{\Lambda}^{theory}/  \rho_{\Lambda}^{observed}\simeq 6.0$

\section{Cosmological constant $\Lambda$ with nonlocal quark
condensate}

   As mentioned above, the expression $\langle 0 \mid : \bar q(x){q}(0):
\mid 0 \rangle = \langle 0 \mid : \bar q(0){q}(0): \mid 0 \rangle
-\frac{x^{2}}{4}\langle 0 \mid : \bar q(0)[ig_{s}\sigma G(0)]{q}(0):
\mid 0 \rangle + \cdots$ does not work except for very small x.
Therefore we shall use the nonlocal quark condensate derived from
the quark distribution function (see Refs.[34,35]). Using the form in
Ref.[35],
\begin{eqnarray}
\label{q(x)q(0)}
   \langle 0 \mid : \bar q(x){q}(0):\mid 0 \rangle&=& g(x^2)
 \langle 0 \mid : \bar q(0){q}(0):\mid 0 \rangle \; ,
\end{eqnarray}
with
\begin{eqnarray}
\label{g(x)}
          g(x)&=& \frac{1}{(1+\lambda^2 x^2/8)^2} \; .
\end{eqnarray}

The value of $\lambda^2$ estimated in Ref.[36] is $\lambda^2 \simeq
0.8 GeV^2$. Using $1/\Lambda_{QCD}$ as the length scale, or
$x^2=(1/0.2 GeV)^2$, one obtains

\begin{eqnarray}
\label{gQCD}
          g(1/\Lambda_{QCD})&=& \frac{1}{2.25^2}=\frac{1}{6.25} \; .
\end{eqnarray}

From this we obtain
\begin{eqnarray}
 \langle 0 \mid : \bar q(x){q}(0):\mid 0 \rangle&=& \frac{1}{6.25}
 \langle 0 \mid : \bar q(0){q}(0):\mid 0 \rangle \; ,
\end{eqnarray}
and
\begin{eqnarray}
\label{rhononlocal}
     \rho_{\Lambda}^{nonlocal\; theory} &\simeq& \frac{1}{6}(3.6 \times 10^{-3}eV)^4
\nonumber \\
           &=& (2.3 \times 10^{-3}eV)^4 \simeq  \rho_{\Lambda}^{observed} \; .
\end{eqnarray}

  Therefore, using the modification of the quark condensate via the nonlocal
condensate, one obtains excellent agreement between the theoretical and
observed cosmological constants.

\section{Summary and concluding remarks}
The cosmological constant $\Lambda$ is an important physical
quantity, which was
introduced by A. Einstein who modified the field equations of his
general theory of relativity to obtain a stationary universe. The
constant has recently been used to explain the observed accelerated
expansion of the universe, but its observational value is about 120
orders of magnitude smaller than the one theoretically computed in
the framework of the currently accepted quantum field theories.
Namely, quantum field theory predicted that vacuum energy density,
$\rho_{\Lambda}$, is of the order of $M_{pl}^{4}$, with $M_{pl} = 1.22 
\times 10^{19} GeV$, which is about 120 
orders of magnitude larger
than the observed value of $\rho_{\Lambda}^{observed} = (2.3\times
10^{-3} eV)^{4}$. This difference is the so called cosmological
constant problem, the worst problem of fine-tuning in physics.

Based on the Veneziano ghost theory of QCD, using a local quark
condensate, we obtained the same result for $\rho_{\Lambda}$
as in Refs[11,12], about a factor of 6 larger than $\rho_{\Lambda}^{observed}$.
 However, $ <0|:\bar{q}(0)q(0):|0>$ is just an
approximation to $ <0|:\bar{q}(x)q(0):|0>$. Using the
nonlocal quark condensate $ <0|:\bar{q}(x)q(0):|0> = g(x)
<0|:\bar{q}(0)q(0):|0>$ we find that the theoretical and observed
values of $\rho_{\Lambda}$ are approximately equal.

The cosmological constant $\Lambda$ is a potentially important
contributor to the dynamical history of the universe. Unlike
ordinary matter, which can clump together or disperse as it evolves,
the vacuum energy is a property of spacetime itself, and is expectd to 
be the same everywhere. If the cosmological costant is the valid model 
of dark energy, a sufficiently large cosmological constant will cause
galaxies and supernovae to accelerate away from us, as has been observed,
in contrast to the
tendency of ordinary forms of energy to slow down the recession of
distant objects.  The value of $\Lambda$ in our present universe is
not well known. A precise determination of
this constant will be one of the primary goals of both theoretical
cosmology and observational cosmology in the near future.

One might doubt the correctness of the Veneziano QCD ghost theory
that we used in this work, since it is an analogue of two-
dimensional theory based on the Schwinger model$^{[18,19]}$,
replacing the vector gauge field by two scalar fields. These scalar
fields have positive and negative norms and cancel with each other,
leaving no trace in the physical subspace. They have small
contribution to the vacuum energy in the curved space. It is known
that the QCD ghost must be an intrinsically vector field in order
for the $U(1)$ problem to be consistently resolved within the
framework of QCD. It seems to be necessary to examine if the
Veneziano mechanism works in terms of the vector ghost fields
instead of the scalar fields used here. However, Ohta and others in
Refs.[36,37,38] have discussed the same problem in  more realistic
four dimensional models, and show that the QCD ghost produces vacuum
energy density $\rho_{\Lambda}$ proportional to the Hubble parameter
which has approximately the right magnitude $\sim (3 \times
10^{-3}eV)^{4}$.

 There is now considerable evidence that the universe began as fireball
in the cosmological vacuum, the so-called "Big Bang", with extremely
high temperature and high energy density. One knows that the quark 
condensate is vastly changed by the QCD phase transition, and this
implies that there is a tempreature (T) dependence of 
$<0|:\bar{q}(x)q(0):|0>$ and $\Lambda$.
 $\Lambda$ is probably dependent on temperature T and momentum $p$
of virtual particles which produce vacuum condensates, as mentioned
above. We can predict the $\Lambda$ dependence on
temperature T and momentum $p$ by solving the temperature dependent
Dyson-Schwinger Equations. In this case, $\Lambda$ is a function of
T and $p$. Such a new study could show the behavior of the $\Lambda$
during the evolution of the universe. This work is under its way and
should be complete soon.

\end{document}